\documentclass[%
 reprint,
 preprintnumbers,
 amsmath,amssymb,
 aps,
]{revtex4-1}
\usepackage{graphicx}% Include figure files
\usepackage{dcolumn}% Align table columns on decimal point
\usepackage{bm}% bold math
\begin{document}

%\preprint{APS/123-QED}
%\preprint{CYCU-HEP-14-12}

%\title{\boldmath Enhanced $B_d^0 \to \mu^+\mu^-$ Decay: What if?}

\title{\boldmath
%$K_L \to \pi^0\nu\bar\nu$ Beyond the Grossman-Nir Bound
Loophole in $K \to \pi\nu\bar\nu$ Search and New Weak Leptonic Forces
}

\author{Kaori Fuyuto$^{a}$, Wei-Shu Hou$^{b}$, and Masaya Kohda$^{c}$}
 \affiliation{
$^{a}$Department of Physics, Nagoya University, Nagoya 464-8602, Japan\\
$^{b}$Department of Physics, National Taiwan University, Taipei 10617, Taiwan\\
$^{c}$Department of Physics, Chung-Yuan Christian University, Chung-Li 32023, Taiwan
}


\begin{abstract}
Weakly interacting $K \to \pi X^0$ emission with $m_{X^0} \cong m_{\pi^0}$
is out of sight of the current $K^+\to \pi^+\nu\bar\nu$ study,
but it can be sensed by the $K_L \to \pi^0\nu\bar\nu$ search.
This evades the usual Grossman-Nir bound of
${\cal B}(K_L \to \pi^0\nu\bar\nu) < 1.4 \times 10^{-9}$, thus
the KOTO experiment is already starting to probe New Physics.
An intriguing possibility is the $Z'$ gauge boson of a weak leptonic force
that couples to $L_\mu - L_\tau$
(the difference between the muon and tauon numbers),
which may explain the long-standing ``muon $g-2$'' anomaly,
but is constrained by $\nu_\mu N \to \nu_\mu N\mu^+\mu^-$ scattering
to $m_{Z'} \lesssim 400$ MeV.
An explicit model for $K\to \pi Z'$ is given, which illustrates
the link between rare kaon and $B\to K\mu^+\mu^-$, $K^{(*)}\nu\bar\nu$ decays.
Complementary to these searches and future lepton experiments,
the LHC might discover the scalar boson $\phi$ responsible for
light $m_{Z'}$ generation via $\phi \to Z'Z' \to 2(\mu^+\mu^-)$. %,
%with $m_{\mu^+\mu^-}$ typically $\sim 300$ MeV in mass.
\begin{description}
% Secondary publications and information retrieval purposes.
\item[PACS numbers]
11.30.Er % CPT and other discrete symmetry
13.20.Eb % Decays of K mesons
13.20.He % Decays of bottom mesons
14.70.Pw % Other gauge bosons
% You may use the \texttt{description} environment to structure your abstract; use the optional argument of the \verb+\item+ command to give the category of each item.
\end{description}
\end{abstract}

\maketitle

%\tableofcontents

%\section{\label{sec:Intro}INTRODUCTION\protect\\}
{\it Introduction}---Despite discovering a
126 GeV scalar boson~\cite{PDG14},
there is anxiety at the Large Hadron Collider (LHC):
no sign of New Physics (NP) has so far emerged.
But NP need not come from high energy.
One long standing hint~\cite{PDG14} is the ``muon $g-2$'' anomaly,
the discrepancy between precision experimental measurement
and Standard Model (SM) calculations.
A new experiment~\cite{Muon_g-2}, Muon g-2,
is under preparation that aims for a factor of four improvement in precision,
with theory efforts to match~\cite{Benayoun:2014tra}.
One attractive NP possibility is a new force that couples to the muon,
for example, gauging~\cite{XG} the difference between the $\mu$ and $\tau$ numbers,
$L_\mu - L_\tau$ (much like gauging electric charge), with an associated gauge boson $Z'$.
The scenario is well protected because, besides the muon, the $Z'$
interacts with only $\tau$s and neutrinos.

The ``muon $g-2$'' anomaly maps out a band in
($m_{Z'}$, $g'$) space~\cite{Altmannshofer:2014cfa}, where $g'$ is the gauge coupling,
and may also explain the so-called ``$P_5^\prime$ anomaly''~\cite{Aaij:2013qta}
in $B^0 \to K^{*0}\mu^+\mu^-$ angular variables.
It was found~\cite{Altmannshofer:2014pba}, however,
that the neutrino trident production or $\nu_\mu N \to \nu_\mu N\mu^+\mu^-$ process
constrains the $Z'$ to be light,
\begin{equation}
m_{Z'} \lesssim 400\ {\rm MeV},
 \label{eq:lightZ'}
\end{equation}
and $g'$ is far weaker than the weak coupling.
If this $Z'$ couples to quarks in some way,
then rare $K$ decays
%--- the longest pursuit since the dawn of particle physics ---
might probe for the existence of this light ${Z'}$.
While contemplating this link, we uncover a loophole in
the usual Grossman-Nir (GN) bound~\cite{Grossman:1997sk},
\begin{align}
{\cal B}(K_L \to \pi^0\nu\bar\nu)
 & < 1.4 \times 10^{-9}.\ \ \ {\rm (``GN\ bound")}
 \label{eq:GNcommon}
\end{align}
%
%${\cal B}(K_L \to \pi^0\nu\bar\nu) < 1.4 \times 10^{-9}$:
Kinematic selection in $K^+\to \pi^+\nu\bar\nu$ search
allows $K^+\to \pi^+Z'$ to go unnoticed, \emph{if} $m_{Z'} \sim m_{\pi^0}$,
but $K_L \to \pi^0Z'$ can be sensed by $K_L \to \pi^0\nu\bar\nu$ search,
%above the bound of Eq.~(\ref{eq:GNcommon})!
%\emph{for the lack of kinematic control!}
thereby the bound of Eq.~(\ref{eq:GNcommon}) is evaded.

Besides pointing out this generic loophole,
in this Letter we give an explicit model (see Fig.~1)
% (see Fig.~1 for how $s\to dZ'$ might be induced)
that also shows how rare kaon and analogous rare $B$ processes are interlinked.
We point out further that the LHC could search for the
scalar boson $\phi$ behind $m_{Z'}$ generation,
via a pair of very light dimuons,
i.e. $\phi \to Z'Z' \to 2[\mu^+\mu^-]$.

\begin{figure}[b!]
%\begin{center}
%\includegraphics[width=60mm,height=40mm]{plots/twidth.pdf}
%\vspace{-3mm}
{
 \includegraphics[width=65mm]{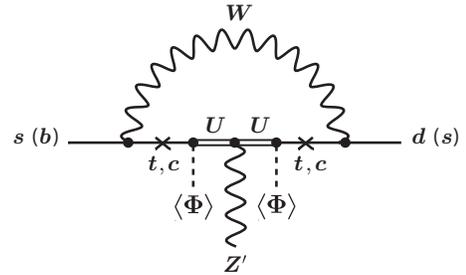}
}
%\end{center}
%\vskip-0.05cm
\caption{
Effective $dsZ'$ ($sbZ'$) coupling, with $Z'$ coupled to a vector-like $U$ quark that
%mixes with $c$, $t$ (an ``$\times$'' flips chirality) and connect
mixes with $c$, $t$ (``$\times$'' flips chirality) and connects
with external $d$-type quarks via a $W$ boson loop.
} \label{fig:s_to_dZ'}
\end{figure}

{\it $K\to \pi\nu\bar\nu$ Search}---The E787/949 experiment~\cite{Artamonov:2009sz}
has measured ${\cal B}(K^+ \to \pi^+\nu\bar\nu) = (1.73^{+1.15}_{-1.05}) \times 10^{-10}$,
which is consistent with SM expectations,
and the NA62~\cite{NA62} experiment aims at
collecting ${\cal O}(100)$ events in next 3 years.
In a similar time frame, the KOTO experiment~\cite{KOTO}
aims at $3\sigma$ measurement of $K_L \to \pi^0\nu\bar\nu$ assuming SM rate.
KOTO has a better chance to uncover NP, because
$K_L \to \pi^0\nu\bar\nu$ decay is intrinsically CP violating (CPV),
and the existing limit~\cite{Ahn:2009gb}
%
%But the current limit %by E391a,
%the predecessor experiment to KOTO, gives
%
\begin{equation}
{\cal B}(K_L \to \pi^0\nu\bar\nu) < 2.6 \times 10^{-8},\ \ \ ({\rm  E391a})
 \label{eq:E391a}
\end{equation}
is weaker. Eq.~(\ref{eq:E391a}) is, however,
far above the bound of Eq.~(\ref{eq:GNcommon}),
which follows from inserting the E787/949 measurement
into the relation~\cite{Grossman:1997sk},
\begin{align}
{\cal B}(K_L \to \pi^0\nu\bar\nu)
 & \lesssim 4.3 \times {\cal B}(K^+ \to \pi^+\nu\bar\nu),
 \label{eq:GN}
\end{align}
where the number 4.3 arises from isospin
and $\tau_{K_L}/\tau_{K^+}$~\cite{Grossman:1997sk}.
This is the origin of the usually \emph{perceived} GN bound,
that KOTO can only probe NP after Eq.~(\ref{eq:GNcommon}) is reached.
But KOTO has suffered a few inadvertent setbacks,
and accumulated just 100 hours of data in 2013.
Though sensitivity comparable to Eq.~(\ref{eq:E391a}) is reached~\cite{KOTO-CKM2014},
there is one event in the signal box,
compared with 0 events for the E391a~\cite{Ahn:2009gb} experiment,
hence KOTO appears to be still far from the bound of Eq.~(\ref{eq:GNcommon}).

{\it Experimental Loophole}---%
The design of experiments have ``accidental'' features
that are akin to the factor of 4.3 in Eq.~(\ref{eq:GN}) being not
just a simple isospin factor.
The E787/949 experiment observes $K^+$ decay at rest,
detects the emitted $\pi^+$, but nothing else.
However, due to the ``brightness'' of
${\cal B}(K^+ \to \pi^+\pi^0) \simeq 21\%$,
the region around $m_{\pi^0}$, i.e. the range of $p_{\pi^+}$
corresponding to $116 \lesssim m_{\rm miss} \lesssim 152$~MeV,
is kinematically excluded.
The region for $m_{\rm miss} > 261$ MeV
is further excluded~\cite{Snowmass} due to $K^+ \to \pi^+\pi\pi$ background.
Although NA62 measures $K^+$ decay in-flight,
the regions of
$100 \lesssim m_{\rm miss} \lesssim 165$ MeV
and $m_{\rm miss} \gtrsim 260$ MeV
are similarly excluded.
%which is a bit tighter. %than E787/949 for sake of background rejection.

A $K_L \to \pi^0\nu\bar\nu$ experiment, however,
cannot do kinematic reconstruction: besides detecting
two photons (assumed as $\pi^0$), it measures ``nothing to nothing''.
The $K_L$ and ``$\pi^0$'' momenta are not known.
The approach is thus to veto everything,
and to learn while pushing down the sensitivity.
However, the $\nu\bar\nu$ being the target,
one cannot veto weakly interacting light particles (WILP).
Thus, for $K \to \pi X^0$ where $X^0$ is \emph{any} WILP that
falls into the missing mass window,
the $K^+$ experiment would be oblivious,
\emph{but the $K_L$ experiment can have a blunt feel!}
Although the GN relation of Eq.~(\ref{eq:GN}) is in no way violated,
the perceived GN bound of Eq.~(\ref{eq:GNcommon}) does not apply.
This is the main and rather simple point of this Letter,
independent of model discussion.
The $X^0$ need not be the leptonic force,
as it simply goes undetected.

The E949 experiment performed a tagged search for
$\pi^0 \to \nu\bar\nu$~\cite{Artamonov:2005cu}
inside the kinematically excluded window %of 120 to 160 MeV,
around $\pi^0$,
giving the 90\% C.L. bound~\cite{Artamonov:2009sz}
\begin{equation}
{\cal B}(K^+ \to \pi^+ X^0) < 5.6 \times 10^{-8},\ \ \ (m_{X^0} = m_{\pi^0})
 \label{eq:K+pi+X}
\end{equation}
which is much weaker than their ${\cal B}(K^+ \to \pi^+\nu\bar\nu)$ bound.
Applying the analog of Eq.~(\ref{eq:GN}) would
imply ${\cal B}(K_L \to \pi^0 X^0) < 2.4 \times 10^{-7}$,
much weaker than the E391a bound of Eq.~(\ref{eq:E391a}).
Hence Eq.~(\ref{eq:E391a}) provides a direct and more stringent
bound on $K_L \to \pi^0 X^0$ than implied by Eq.~(\ref{eq:K+pi+X}),
which illustrates our main point.

We now give an explicit $Z'$ model
to illustrate the potential impact of a $K_L \to \pi^0 X^0$ discovery.

%An explicit $Z'$ model is now given as existence proof.

%\section{Explicit Model}
%\vskip0.2cm

{\it Explicit Model}---We were interested in
$t\to cZ'$ decay %involving a very \emph{light} $Z'$,
in the model of Ref.~\cite{Altmannshofer:2014cfa},
where tree level $sbZ'$ and $ctZ'$ couplings
are generated through mixing of SM quarks with vector-like
doublet $Q$ and singlet $D$, $U$ quarks. %(obvious notation).
With the $Z'$ boson of gauged $L_\mu - L_\tau$
solution to muon $g-2$ anomaly constrained~\cite{Altmannshofer:2014pba}
by neutrino trident production to be light, Eq.~(\ref{eq:lightZ'}),
one is motivated to study rare $K$ decay,
%
%A \emph{heavy} $Z'$ boson
%could account for the ``$P_5^\prime$ anomaly''~\cite{Aaij:2013qta}
%in $B^0 \to K^{*0}\mu^+\mu^-$ angular variables,
%but cannot be our light $Z'$.
%The model can operate even if one does not consider the $P_5'$ anomaly.
but the model can still be applied.

For $s\to d$ transitions,
mixing in the down-type sector would become too fine-tuned,
hence setting them to zero is reasonable,
and we consider only mixing of up-type quarks with $U$,
which is less constrained.
This can be achieved, e.g. by introducing a $Z_2$ symmetry
under which $Q$ and $D$ are odd while $U$ and other fields
are even~\cite{zeros}.
%
%In the model of Ref.~\cite{Altmannshofer:2014cfa} with
%no mixing in down-type quarks,
Diagrams like {Fig.~1} can start from a $U$ and $t$, $c$ mixing core
where the $Z'$ is emitted, and dressed up \emph{with assistance from SM}
into a loop-induced $s\to dZ'$ (or $b\to sZ'$) transition.
The loop is finite because tree level down-type mixing is set to zero.

It is intriguing that with reasonable $Uc$ and $Ut$ mixing parameters
(but with $Uu$ mixing set to zero),
loop diagrams as in Fig.~1 bring the $s\to d$ transition into
current experimental sensitivities.
To introduce our subsequent notation, note that 
the vector-like quark $U$ in Fig.~1 carries the extra U(1)$'$ charge 
hence emits the $Z'$ boson, while it mixes with right-handed up-type quarks $i = c$, $t$
through a ``Yukawa coupling'' $Y_{Ui}$ 
to an exotic scalar field $\Phi$ with U(1)$'$ charge
(and $\langle \Phi\rangle = v_\phi/\sqrt{2}$ generates $m_{Z'}$).
For more details, see Ref.~\cite{FHK2}.

%
%Using the model of Ref.~\cite{Altmannshofer:2014cfa}
%with only the $U$ quark mixing with $c$ and $t$,
The effective $\bar d_L\gamma^\mu s_L Z_\mu^\prime$ coupling~\cite{FHK2}
has coefficient
%that generates $K \to \pi Z'$ decay,
%
\begin{equation}
g_{ds} =\frac{g'v_\phi^2}{32\pi^2v^2}\left[
 c_{cc}f_{cc} + (c_{tc}+c_{ct})f_{ct} + c_{tt}f_{tt}
 \right]
 , %+{\rm h.c.}
 \label{eq:dsZ'}
\end{equation}
%
%in the 't Hooft--Feynman gauge, %including Goldstone boson diagrams,
where %$g'$ is the $Z'$ gauge coupling,
%$v_\phi$ is the extra U(1)$'$ breaking scale,
$c_{ij} = V_{is}V_{jd}^* Y_{Ui}Y_{Uj}^*{m_im_j}/{m_U^2}$,
%$Y_{Ui}$ are Yukawa couplings, 
and
\begin{align}
%f_{cc} &= 4\log\frac{m_W^2}{m_c^2} + \log\frac{m_U^2}{m_W^2}-3, \notag \\
f_{ct} &= 1 + \log\frac{m_U^2}{m_t^2}
            + \frac{3m_W^2}{m_t^2-m_W^2}\log\frac{m_t^2}{m_W^2}, \notag \\
f_{tt} &= \frac{3m_W^2}{m_t^2-m_W^2}\left( 1-\frac{m_W^2}{m_t^2-m_W^2}
          \log\frac{m_t^2}{m_W^2}\right) + \log\frac{m_U^2}{m_t^2}, \notag
\end{align}
with $f_{cc}$ obtainable from $f_{tt}$ in $m_t^2 \ll m_W^2$ limit.
These expressions are in the large $m_U$ limit,
though we use exact one-loop expressions (see Ref.~\cite{FHK2})
in our numerics.
Note that $c_{ct} \neq c_{tc}$, and
$c_{ij}$ are complex, even for real $Y_{Ui}$.

\begin{figure*}[t!]
%\begin{center}
%\includegraphics[width=60mm,height=40mm]{plots/twidth.pdf}
%\vspace{30mm}
{\includegraphics[width=72mm]{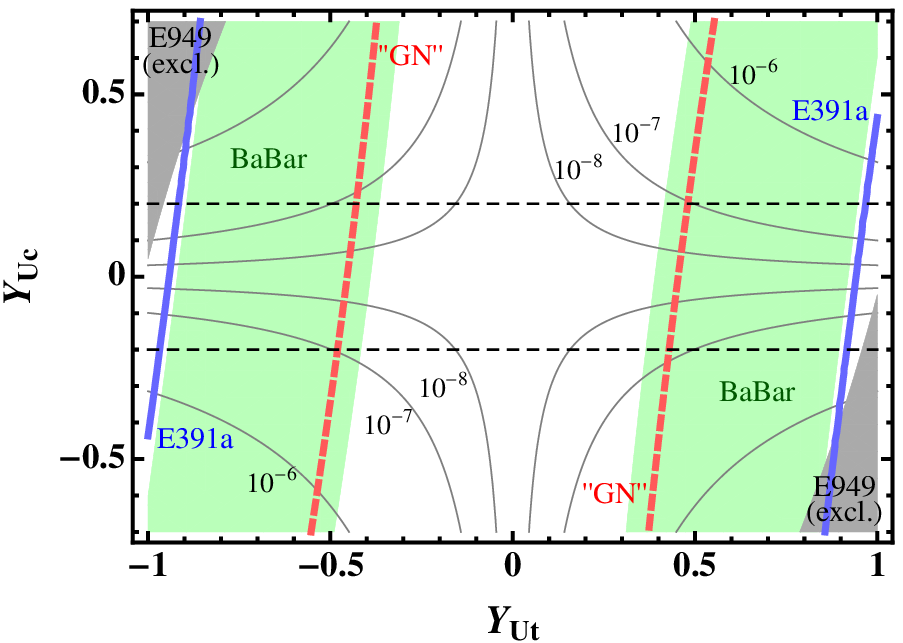} \hskip0.5cm
 \includegraphics[width=72mm]{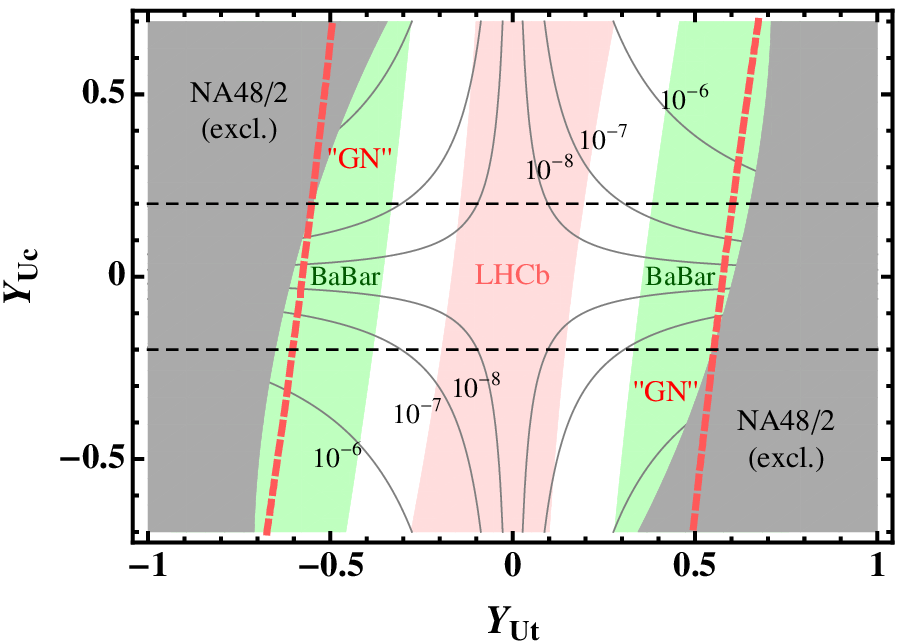}
}
%\end{center}
%\vskip3.5cm
\vskip0.05cm
\caption{
[left]
For $m_{Z^\prime}=135$ MeV ($Z^\prime \to \nu\bar\nu$ 100\%),
bounds for $\mathcal B(K^+\to\pi^+ Z^\prime) < 5.6\times 10^{-8}$
(dark grey exclusion region) and $\mathcal B(K_L \to \pi^0 Z^\prime)< 2.6\times 10^{-8}$
(blue solid) on the $Y_{Uc}$-$Y_{Ut}$ plane.
[right]
For $m_{Z^\prime}=285$ MeV ($Z^\prime \to\nu\bar\nu$ 54\%),
bounds for $\mathcal B(K^+ \to \pi^+ Z^\prime)
\mathcal B(Z^\prime \to \mu^+\mu^-) < 2.1\times 10^{-9}$ (dark grey exclusion region)
and $\mathcal B(B^+ \to K^+Z^\prime)\mathcal B(Z^\prime\to \mu^+\mu^-)
<2.0\times 10^{-8}$ (pink allowed region) on the $Y_{Uc}$-$Y_{Ut}$ plane.
In both panels, we give the usual ``GN bound'' of $\mathcal B(K_L\to \pi^0 Z^\prime)
\mathcal B(Z^\prime \to \nu\bar\nu) < 1.4\times 10^{-9}$ (red dashed) and
$2\sigma$ range for $\mathcal B(B^+\to  K^+Z^\prime)\mathcal B(Z^\prime\to\nu\bar\nu)
=(0.35^{+0.6}_{-0.15})\times 10^{-5}$ (light green allowed region).
The horizontal lines mark reasonable $Y_{Uc}$ range, and in the backdrop
we plot $\mathcal B(t\to cZ^\prime)$ contours.
}
\end{figure*}

The branching ratio for $K^+\to\pi^+ Z^\prime$ is given by
\begin{align}
&{\cal B}(K^+\to \pi^+ Z^\prime) \notag \\
&=\frac{m_{K^+}}{\Gamma_{K^+}} \frac{|g_{ds}|^2}{64\pi \hat{m}_{Z'}^2}
 \lambda^{{3}/{2}}\left(1,\hat{m}_{\pi^+}^2,\hat{m}_{Z^\prime}^2\right)
 \left[f_+^{K\pi}\left(m_{Z^\prime}^2\right)\right]^2,
 \label{eq:KpiZ'}
\end{align}
where $\lambda(x,\,y,\,z) \equiv x^2 + y^2 + z^2 -2(xy+yz+zx)$,
$\hat{m} \equiv m/m_{K^+}$, and $f_+^{K\pi}$ is a form factor.
The formula for $K_L\to \pi^0 Z'$ is analogous,
with $|g_{ds}|$ replaced by ${\rm Im}\,g_{ds}$.
Taking $f_+^{K\pi}$ values from Ref.~\cite{Mescia:2007kn},
we plot in {Fig.~2[left]}
the bound of Eq.~(\ref{eq:K+pi+X}) for $K^+ \to \pi^+ Z'\vert_{m_{Z'} = m_{\pi^0}}$
in the $Y_{Uc}$--$Y_{Ut}$ (treated as real) plane.
We have taken $g' \sim 10^{-3}$ as fixed~\cite{Altmannshofer:2014pba}
by muon $g-2$ excess and neutrino trident bound,
and $m_U = 2$ TeV, $v_\phi = 135$ GeV.
We also plot $K_L \to \pi^0 Z'$ assuming the E391a bound of Eq.~(\ref{eq:E391a}),
which turns out comparable.
%, even though Eq.~(\ref{eq:K+pi+X}) probably should not be applied.
But if we apply Eq.~(\ref{eq:GNcommon})
as a bound on $K_L \to \pi^0 Z'$ (``GN" in Fig.~2), it would be much more
stringent than the direct bound of Eq.~(\ref{eq:E391a}).
We have argued, however, that this application of ``GN bound'' is
incorrect for the present case.
Hence, the region between Eq.~(\ref{eq:E391a}) and
Eq.~(\ref{eq:GNcommon}) is fair game for discovery!
Note that $K_L \to \pi^0 Z'$ is sensitive to the
imaginary part of $dsZ'$ coupling in Eq.~(\ref{eq:dsZ'}),
%($K^+ \to \pi^+ Z'$ depends on the absolute value),
hence probes also extra CPV phases arising from $Y_{Uc}$ and $Y_{Ut}$.
Other curves and regions in Fig.~2[left] would be explained shortly.

For the $m_{\rm miss} > 260$ MeV exclusion zone for
$K^+\to\pi^+\nu\bar\nu$, $Z' \to \mu^+\mu^-$ decay is allowed.
We find ~\cite{FHK2} that the $K^+ \to \pi^+\mu^+\mu^-$ data
by the NA48/2 experiment~\cite{Batley:2011zz} permits
a ``best possible spike'' at $m_{\mu\mu} \simeq 285$ MeV,
with $\delta{\cal B}(K^+ \to \pi^+\mu\mu)$ up to $2.1\times 10^{-9}$ in strength.
This is plotted (dark grey exclusion region) in {Fig.~2[right]}
%on $Y_{Uc}$--$Y_{Ut}$ plane,
and is as stringent as the ``GN bound'' of Eq.~(\ref{eq:GNcommon}),
hence much more stringent than Eq.~(\ref{eq:E391a}).
The model parameters are $g' = 1.3 \times 10^{-3}$,
$m_U = 2$ TeV and $v_\phi \simeq 219$ GeV.

We have shown that KOTO is already starting to probe NP.
If a genuine excess appears above the perceived ``GN bound'' of Eq.~(\ref{eq:GNcommon}),
the likely explanation would be an unobserved recoil $X^0$ particle
in the ``$\pi^0$ exclusion window'' of $K^+ \to \pi^+\nu\bar\nu$ search.
Note that the bound of Eq.~(\ref{eq:GNcommon})
cannot improve by much, even as NA62 accumulates data,
unless ${\cal B}(K^+ \to \pi^+\nu\bar\nu)$ is found
to be below SM expectation.
If KOTO pushes down to this bound of Eq.~(\ref{eq:GNcommon}) without discovery,
then NA62 should scan above 260 MeV for dimuon peaks.
It could also push the bound on $\pi^0\to\nu\bar\nu$~\cite{Artamonov:2005cu}
in the $m_{\pi^0}$ exclusion window,
and extend the study of E787/949 for $K^+ \to \pi^+X^0$
(see Fig.~18 of Ref.~\cite{Artamonov:2009sz} and discussion).
With sufficient statistics, one might still uncover
peaking events in $m_{\rm miss}$.

We remark that, for both cases of discussion, we have checked that
the benchmark parameters satisfy the kaon mixing constraint of
Eq.~(11) in Ref.~\cite{He:2005we}.

%
%We turn now to other
%implications of our model.

%\section{Further Model Implications}
%\vskip0.2cm

{\it Further Model Implications}---We have kept $Uc$ and $Ut$ mixings but set the
mixing of heavy vector-like quarks with down-type quarks (as well as $u$) to zero.
But Fig.~1 generates $sbZ'$ couplings alongside $dsZ'$ couplings
by $W$ exchange in the loop.
This brings in rare $B$ decays,
where the LHCb experiment has demonstrated its prowess recently,
while Belle II is under construction.
The formulas are analogous to Eqs.~(\ref{eq:dsZ'}) and (\ref{eq:KpiZ'}).

For the $m_{Z'} = 285$ MeV case that we have just illustrated,
$Z' \to \mu^+\mu^-$ and $\nu\bar\nu$ rates are comparable,
and the decay is prompt.
Thus, it can show up in $B\to K^{(*)}\mu\mu$ decay
with very low $m_{\mu\mu}$.
The LHCb experiment has updated differential rates~\cite{Aaij:2014pli} for
$B \to K^{+,0}\mu\mu$ and $K^{*+}\mu\mu$ decays
to 3 fb$^{-1}$, or full Run 1 dataset.
The $B^0 \to K^{*0}\mu\mu$ decay,
relevant for the $P_5^\prime$ anomaly,
has yet to be updated from 1 fb$^{-1}$ data~\cite{Aaij:2013qta}.
But perhaps influenced by the latter,
Ref.~~\cite{Aaij:2014pli} starts at $q^2 \equiv m_{\mu\mu}^2 > 0.1$ GeV$^2$,
or $m_{\mu\mu} \gtrsim 316$ MeV,
which covers only half the region of $m_{\mu\mu}$ allowed by
Eq.~(\ref{eq:lightZ'}) above the dimuon threshold.

The 1 fb$^{-1}$ paper for $B^+ \to K^+\mu\mu$~\cite{Aaij:2012vr},
however, does go down to $q^2 = 0.05$ GeV$^2$, or $m_{\mu\mu} = 224$ MeV,
hence can be compared with our $m_{Z'} = 285$ MeV case.
Interestingly, in the lowest $0.05 < q^2 < 2.00\ {\rm GeV}^2$ bin,
there is a mild excess above the mean for $1.00 < q^2 < 6.00\ {\rm GeV}^2$.
Treating experimental error at the 2$\sigma$ level,
our estimate~\cite{FHK2} for this excess is $\sim 2\times 10^{-8}$.
If we attribute this all to the presence of $B^+\to K^+Z'[\to \mu^+\mu^-]$,
then scaling by ${\cal B}(Z'\to \mu^+\mu^-) \simeq 46\%$,
this implies $B^+\to K^+Z'$ at {4.4} $\times 10^{-8}$ level.
Using form factors of Ref.~\cite{Ball:2004ye},
we plot this constraint in {Fig.~2[right]},
which is stronger than our estimate of the NA48/2 bound.
Actually, there also seems to be some excess
in the first $0.1 < q^2 < 0.98\ {\rm GeV}^2$ bin
for $B \to K^+\mu\mu$ in the full 3 fb$^{-1}$ dataset~\cite{Aaij:2013qta},
hence the $Z'$ could be above 316 MeV.
We urge LHCb to refine their analysis,
optimize binning to $q^2$ resolution,
and extend a spike search down to 0.045 GeV$^2$.

The $B^0 \to K^0\mu\mu$ modes has less statistics,
while $B \to K^{*}\mu\mu$ would have a low $q^2$ photon peak,
making interpretation more difficult.
%
%Note that our estimate based on 1 fb$^{-1}$ LHCb data is stronger than NA48/2, perhaps because the latter is not quite a high statistics study.
Note that our estimate based on LHCb data is stronger than NA48/2,
even though the former is only based on the 1 fb$^{-1}$ dataset.
However, $s\to d$ and $b\to s$ processes may or may not be
correlated as in our model.
So, when KOTO reaches the usual ``GN bound'', NA62 should still
conduct a spike search above $m_{\mu\mu} > 260$ MeV.
We note in passing that the Belle experiment has conducted
$B^0 \to K^{*0}X^0$ search~\cite{Hyun:2010an}
for light $X^0 \to \mu^+\mu^-$,
and the bound is roughly $5 \times 10^{-8}$
for $m_{X^0} \simeq 285$ MeV. %, if we assume 50\% dimuon branching.
We suggest Belle (and BaBar), however, to conduct the search for
$B \to K+X^0[\to \mu^+\mu^-]$ to avoid the photon peak.

Like our illustration in Fig.~2[left], if $m_{Z'}$ falls into
the ``$\pi^0$ blind spot'', NA62 would be oblivious, and so would LHCb.
Fortunately, because ${\cal B}(B\to K\pi^0) \ll {\cal B}(K\to \pi\pi^0)$,
the (super-)B factories can crosscheck in
the $B \to K^{(*)}\nu\bar\nu$ modes, where there is no photon peak.
The BaBar experiment has lead the way by conducting
a binned $m_{\nu\bar\nu}^2$ search~\cite{Lees:2013kla},
where the lowest $s_B \equiv m_{\nu\bar\nu}^2/m_B^2 < 0.1$ bin
for both the $B^+ \to K^+\nu\bar\nu$ and $B^0 \to K^{*0}\nu\bar\nu$ modes
show some excess, which drives a \emph{lower bound} for the $K^+\nu\bar\nu$ mode.
From Fig.~6 of Ref.~\cite{Lees:2013kla}, we estimate
${\cal B}(B^+ \to K^+\nu\bar\nu) = (0.35^{+0.6}_{-0.15}) \times 10^{-5}$
in this bin, and plot the 2$\sigma$ range in Fig.~2[left].
The result is stronger than the kaon modes,
and the allowed region extends to the usual ``GN bound''.
On the other hand, for the $m_{Z'} = 285$ MeV example
where $Z'\to \mu^+\mu^-$ is also allowed,
plotting the BaBar result in Fig.~2[right] shows some tension with
our LHCb 1 fb$^{-1}$ estimate for $B^+\to K^+Z'[\to \mu^+\mu^-]$,
with the latter most stringent.
Our estimates are, however, rudimentary and for illustration only.
It would be better done by the experiments.

In this vein, although Belle lead the way in
$B^+ \to K^+\nu\bar\nu$ search~\cite{Chen:2007zk},
its follow-up paper~\cite{Lutz:2013ftz}
just added 40\% data but followed the same analysis,
including a cut on high $p_{K^+}$ for sake of rejecting $B\to K^*\gamma$,
which precisely cuts out the $B\to K^{(*)}Z'$ possibility.
We urge Belle to conduct a binned $m_{\nu\bar\nu}^2$ study
and optimize the binning according to resolution.
It should also practice optimizing the $m_{\nu\bar\nu}^2$ or recoil mass
resolution with the full B-tag method, towards a future Belle~II search.

%\section{Discussion and Conclusion}
%\vskip0.2cm

{\it Discussion and Conclusion}---We have given the
branching ratio ${\cal B}(t\to cZ')$ in the backdrop of Fig.~2,
and have drawn $|Y_{Uc}| < 0.2$ (arbitrarily chosen) bands 
to indicate that $|Y_{Uc}|$ should not be too large,
while $|Y_{Uc}| < |Y_{Ut}|$ should hold in general
(further discussion is given in Ref.~\cite{FHK2}).
%Although tree level in nature, 
We find ${\cal B}(t\to cZ') \lesssim 10^{-7}$
for $|Y_{Uc}| < 0.2$,
but the rate can certainly be larger if one
considers general $|Y_{Uc}|$ values.
Thus, %is in good part due to rather weak $g'$, making 
given that $Z' \to \mu^+\mu^-$ at $\sim 50\%$ for
$Z'$ above the dimuon threshold, 
$t\to cZ'$ %seemingly inaccessible 
should be searched for at the LHC,
while the rare $t\to cZ'$ case could perhaps drive 
a 100 TeV pp collider study for a future ``top factory''.
%But there is one hope, not related to flavor physics.

With spontaneous $L_\mu - L_\tau$ symmetry breaking
but $Z'$ light because of very weak gauge coupling,
the $v_\phi$ scale is not too different from $v$ of SM.
The mass of the exotic scalar $\phi$ is quite arbitrary as
the self coupling is unknown, but should be at the weak scale.
However, the $U$ quark mixes with the $c$ and $t$ quarks,
which generates \emph{effective $gg\phi$ coupling},
while $\phi$ predominantly decays via a $Z'Z'$ pair.
This motivates a search for the light $Z'$ boson at the LHC,
which can potentially uncover the associated $\phi$ boson,
\emph{independent of rare $K$ and $B$ studies}.

Our investigation~\cite{FHK2} shows that
$\phi$ search is accessible at the LHC for
the example of a 285 MeV $Z'$, where the signature is
$(gg \to)~\phi \to Z'Z' \to [\mu^+\mu^-] [\mu^+\mu^-]$
with brackets indicating low dimuon mass.
The $Z'$ decay is prompt.
Interestingly,
%after an initial study~\cite{Chatrchyan:2012am} with 2011 data,
the CMS experiment conducted a search~\cite{CMS:2013lea} with 2012 data
that can be applied to $\phi \to Z'Z' \to (\mu^+\mu^-)\, (\mu^+\mu^-)$,
where \emph{one event was found at low dimuon pair mass}.
The two dimuon pairs have masses $\sim 200$, 300 MeV, respectively,
which is right on the spot.
It is too early to tell, but with Run 2 to start in 2015,
this study should be carefully watched, and vigorously pursued.
Note that the $U$ quark, with mass in TeV range, can also be searched for.

For the original motivation, muon $g-2$ is pursued by
the E989 or Muon g-2 experiment~\cite{Muon_g-2}, while
neutrino trident production can~\cite{Altmannshofer:2014pba} be
covered by the LBNE experiment~\cite{LBNE}.
Although the schedule is yet uncertain for these two pursuits at Fermilab,
we have shown that the next few years could see
major progress on related issues, ranging from
rare kaon decays (KOTO/NA62), rare B decays (LHCb/Belle(II)),
and perhaps the LHC.

In conclusion,
we point out a loophole in the
experimental setup when comparing
$K^+ \to \pi^+\nu\bar\nu$ and $K_L \to \pi^0\nu\bar\nu$ search,
and find that the KOTO experiment is already starting to
explore New Physics territory, while the
commonly perceived ``Grossman-Nir bound'' may not apply.
Although the mass range for
weakly interacting light particle emission
is a bit restricted, our explicit model
illustrates the potential wide-ranging impact of
discovering ${\cal B}(K_L \to \pi^0\nu\bar\nu) \gtrsim 1.4 \times 10^{-9}$.
Conversely, many measurements at B factories and the LHC could
uncover correlated phenomena, which could
shed light on what may be behind the muon $g-2$ anomaly.

\vskip0.3cm
\noindent{\bf Acknowledgement}.
KF is supported by Nagoya University Program for Leading Graduate Schools,
``Leadership Development Program for Space Exploration and Research'' (N01) by JSPS.
WSH is supported by the Academic Summit grant MOST 103-2745-M-002-001-ASP,
as well as by grant NTU-EPR-103R8915.
MK is supported under NSC 102-2112-M-033-007-MY3.
WSH thanks T. Blake, P.~Chang, K.-F.~Chen, Y.B.~Hsiung, M. Pepe-Altarelli
and T.~Yamanaka for discussions,
and KF thanks the NTUHEP group for hospitality during exchange visits.

%%%%%

%%%%%


\begin{thebibliography}{99}
\bibitem{zeros}
 This can also be achieved by $Z'$ charge assignment.
\end{thebibliography}
\end{document}